## A Luminescent Organic Triradical as an Optically Addressable Qudit

Kueyoung E. Kim[1,†], Avik Bhanja[2,†], Alistair Inglis[3,†], William K. Myers[4], Sam L. Bayliss[3,*], Richard H. Friend[1,*], Petri Murto[2,*]

1 The Cavendish Laboratory, University of Cambridge, Cambridge, UK.
2 Department of Chemistry and Materials Science, Aalto University, Espoo, Finland.
3 James Watt School of Engineering, University of Glasgow, Glasgow, UK.
4 Centre for Advanced ESR, Department of Chemistry, University of Oxford, Oxford, UK.

† Equal Contribution
* Corresponding Author

### Abstract

Luminescent molecular trityl radicals provide options for optically-addressable electron spins and are promising for quantum sensing. Here, we present a luminescent triradical comprising three trityl groups covalently attached to a central phenyl ring. We find two distinct photoluminescence bands at 633 nm and 690 nm with different temporal evolutions, which we attribute to excited quartet and doublet states respectively. The ground state quartet is 1 meV (11.5K) lower in energy than the doublet, resulting in near-equal population of doublets and quartets at room temperature. We demonstrate optically initialized spin polarization, coherent control of the ground state quartet, and optical readout. Potential for optical quantum sensing is shown via optically detected magnetic resonance at room temperature and magneto-photoluminescence up to 3% at 5 mT. This material is a candidate optically addressable molecular qudit, providing opportunities to leverage larger state spaces than qubit systems.

### Introduction

Molecular qubits with spin-optical interfaces are promising platforms for quantum sensing.[1,2] Compared to defects in semiconductors,[3–6] molecules provide a high degree of synthetic tunability and precise control over spatial localization which is valuable for sensing applications. To resolve the small signals present in biology and other contexts, qubits should be designed to maximize sensitivity to the relevant external quantity.[7] In turn, qu*dits*—which are qubit analogs with more than two states—are appealing due to their potential to achieve functionality not possible with qu*bits*. The presence of more states can unlock entangling operations,[8,9] multi-quantum sensing protocols,[10] and quantum error correction,[11] all of which can enhance sensitivity past what is possible for qubits.

Several molecular qudit candidates have been reported with potential for quantum sensing. Organometallic complexes can be designed to have a spin-optical interface with high-spin metal centers.[12–14] However, the inclusion of heavy atoms leads to significant spin-orbit coupling which often causes prohibitively short coherence times at room temperature.[8] In contrast, organic molecules contain only light atoms which enables long coherence times even at room temperature.[15–19] Singlet exciton fission materials and radical-chromophore dyads both access high-spin states upon photoexcitation.[20–25] High-spin states can be optically initialized and controlled with microwaves and magnetic fields, demonstrating their utility for quantum sensing. However, both singlet fission materials and radical-chromophore dyads store spin information in excited electronic states, requiring careful engineering of excited state lifetimes and optical read-out mechanisms to recover accumulated information during a sensing protocol.

In contrast, stable organic radicals can store spin information in the ground state. Our group has recently reported organic diradicals based on meta-linked trityl (TTM) radicals with a phenyl or pyridyl linker to have differentiated photoluminescence in separate spin singlet and triplet manifolds, both of which show radiatively-allowed transitions to their respective ground states for optical readout.[15,16,26] Considering the opportunities of qudits over qubits, a natural extension is then to add a third spin to form a luminescent triradical. While recent advances have produced air-stable carbon-centered quartets and triazatriangulenes with large doublet-quartet energy gaps (0.14 eV),[27–30] these systems are generally non-luminescent. However, we note that Lambert and coworkers recently reported a luminescent triradical based on three pyridyldiphenylmethyl radicals coupled to a central triphenylamine,[31] demonstrating photoluminescence in separate doublet and quartet manifolds.

Here, we present a luminescent triradical composed of three TTMs coupled around a central phenyl ring. This phenyl group preserves the 'alternant hydrocarbon' character of the TTM units.[32] This leads to a small dipolar coupling between spins which subsequently allows the triradical to respond to small magnetic fields (5 mT) and low microwave frequencies (~100 MHz).

## Results

**Material Design and Synthesis.** The triradical $PhR_3$ was built by connecting three αHTTM units to a central phenyl linker via Suzuki–Miyaura coupling,[33] followed by complete deprotonation and three-electron oxidation (**Fig. 1a**). Detailed synthetic protocols, structural characterization and systematic chemical names are described in the Supplementary Information (**Figs. S.I.1–S.I.7**). The three radicals in $PhR_3$ are connected through an odd number of linker atoms (meta-link) that sets up a 3-fold ferromagnetic spin interaction.[16,26,34] The phenyl linker preserves the radical spins' localization while maintaining a small exchange coupling (vide infra). The development of high-spin organic qudits requires accurate control over electronic communication between the radical centers precisely in a 'star-like' framework.[35] We opted to use mesityl substituents at the para position in the radicals for added solubility and steric bulk. Similar aryl substitution has also been shown to brighten trityl radicals.[36,37] Isolated yields were ca. 60% for both Suzuki-Miyaura coupling and radical conversion which are consistent with our previous TTM derivatives.[38] There is no evidence for a singly or doubly deprotonated impurity, as shown by the absence of NMR signal (**Fig. S.I.5**).[16]

The cyclic voltammogram of $PhR_3$ (**Fig. S.I.7**) shows a single, well-resolved reversible redox event for both reduction and oxidation steps. The single reduction step at -0.98 V vs. Fc /Fc (peak separation, $\Delta E_{PP}$ = 120 mV) and the oxidation step at 0.73 V vs. Fc /Fc ($\Delta E_{PP}$ = 120 mV) suggesting all three radical centers are reduced to their anionic forms or oxidized to their cationic forms as a single electrochemical process, consistent with the symmetric coupling of the three radicals.

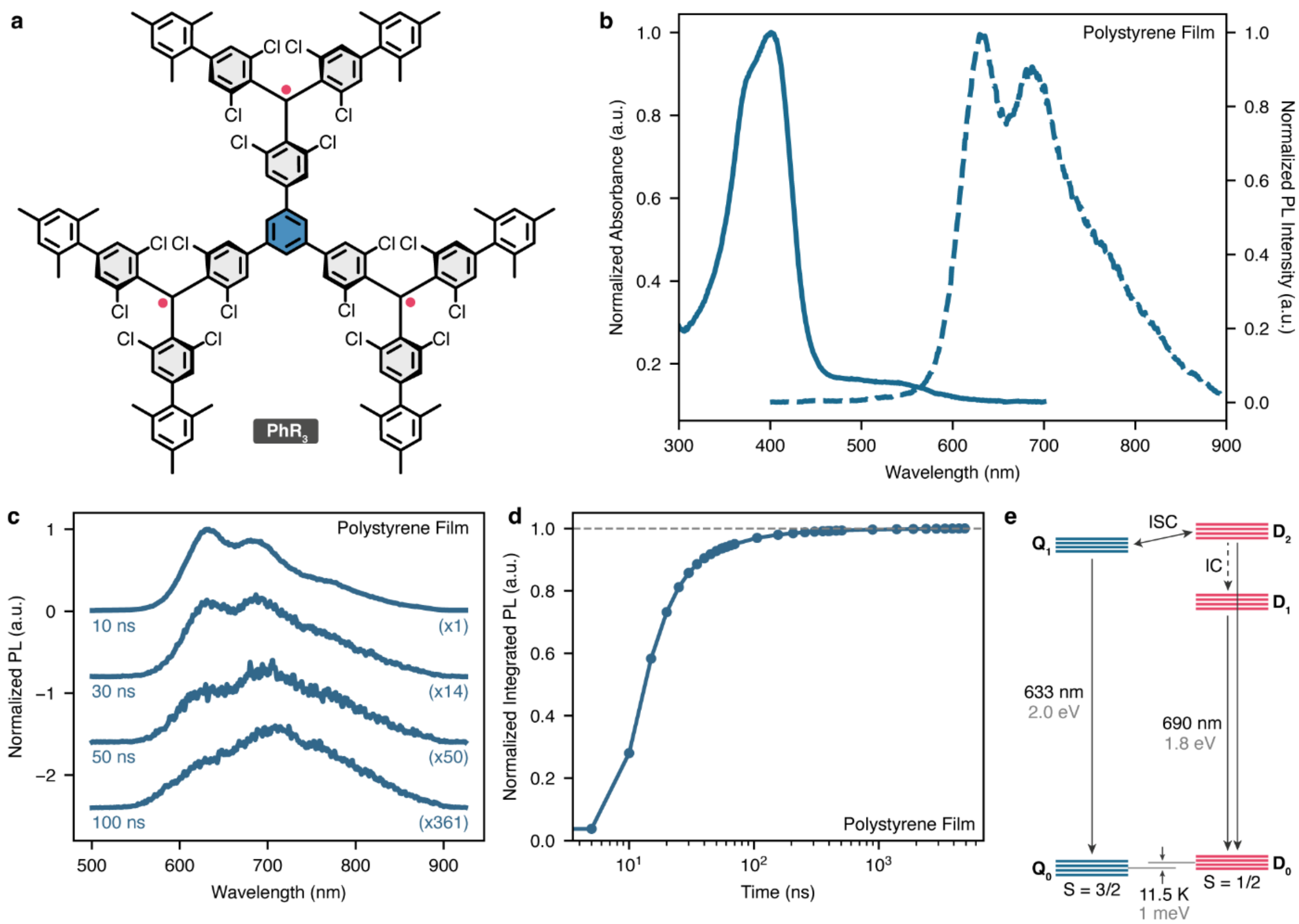

**Figure 1. A luminescent triradical with energetically differentiated doublet and quartet spin states. (a)** Structure of the triradical $PhR_3$. **(b)** Normalized steady-state absorption and PL spectra of a 0.05 wt% $PhR_3$ in polystyrene film. **(c)** Transient PL spectra of $PhR_3$ following photoexcitation at 405 nm. Spectra are normalized for comparison; scale factors are shown on the right. Laser fluence was ~50 nJ/cm$^2$. **(d)** Cumulative PL of $PhR_3$ following 405 nm excitation over time. Integrated PL is calculated from the transient PL spectra shown in (c). **(e)** Proposed energy level diagram for the triradical. In the ground state, the doublet and quartet manifolds are split by an exchange energy of 11.5K (1 meV). The quartet and doublet manifolds are connected by intersystem crossing (ISC). Within the doublet manifold, there is ultrafast (~ps) internal conversion (IC) from the $D_2$ state with local excitation character to the zwitterionic $D_1$ state. Excited doublets primarily emit at 690 nm from the $D_1$ state, while quartets emit at 633 nm. All data is collected on a 0.05 wt% $PhR_3$ in polystyrene film.

**Photophysical and Magnetic Characterization.** The steady-state absorption and photoluminescence (PL) spectra of a dilute 0.05 wt% $PhR_3$ polystyrene film are shown in Fig. 1b. The triradical has a photoluminescence quantum efficiency of 0.6% in polystyrene films (Fig. S.II.1). The PL spectrum contains two distinct peaks at 633 and 690 nm, with a clear shoulder extending into the deep red. Transient PL spectra (Fig. 1c) show that these two peaks originate from distinct electronic states, with the 633 nm band decaying faster than the 690 nm band. Over 95% of the PL occurs within the first 100 ns following photoexcitation (**Fig. 1d**), with the delayed PL primarily originating from the 690 nm band as shown in **Figure 1c**. Time-correlated single photon counting (TCSPC) confirms biexponential kinetics with lifetimes of 7.9 ± 0.1 ns and 92.6 ± 1.1 ns (**Fig. S.II.2**), which we attribute to the 633 and 690 nm PL features respectively.

To probe the character of the excited electronic states, we perform transient absorption (TA) on a 10 $\mu$M solution of $PhR_3$ in toluene (**Fig. 2a**). We observe a sharp excited state absorption (ESA) at 520 nm and broad ESA features at 650 and >800 nm. In line with previous reports, we assign the 520 nm feature to localized absorption on a single TTM site,[39–41] while the lower energy ESA features stem from a zwitterionic state within the doublet manifold.[41–43] The >800 nm ESA forms over ~50 ps (**Fig. 2b**), which we take to be the time for internal conversion from a locally-excited doublet exciton ($D_2$) to the zwitterionic exciton ($D_1$). The formation of the zwitterionic species is supported by spectroelectrochemistry on $PhR_3$ in degassed dichloromethane, shown in **Figure 2c**. Absorption spectra recorded at different applied potentials during the reduction and oxidation of $PhR_3$ are reported in the Supplementary Information (**Fig. S.II.3**). The dominant absorption band of neutral $PhR_3$ at 401 nm decreases when reduced, while two sharp bands at 494 and 600 nm emerge corresponding to the anionic species. Similarly, a broad absorption band centered at 750 nm appears following oxidation, consistent with formation of the cationic species. Notably, the appearance of absorptions at 600 and 750 nm upon reduction and oxidation of $PhR_3$ lends support to the assignment of the low energy ESA features in the TA to a zwitterionic species.

These results are summarized by the energy level diagram shown in **Figure 1e**. In line with our previous report on meta-phenyl linked diradicals,[26] we assign the 633 nm band to PL within the quartet manifold ($Q_1$ to $Q_0$) and the 690 nm band both to a vibrational overtone of the $Q_1$ state and also emission within the doublet manifold ($D_1$ to $D_0$). We expect that only the doublet can access the lower energy zwitterionic state ($D_1$) either through resonant excitation or ultrafast relaxation from a higher-lying exciton with local excitation character ($D_2$), resulting in spin-controlled PL.[26,41–43] The ultrafast (~ps) timescale for this internal conversion is important for optical spin-polarization, discussed below.

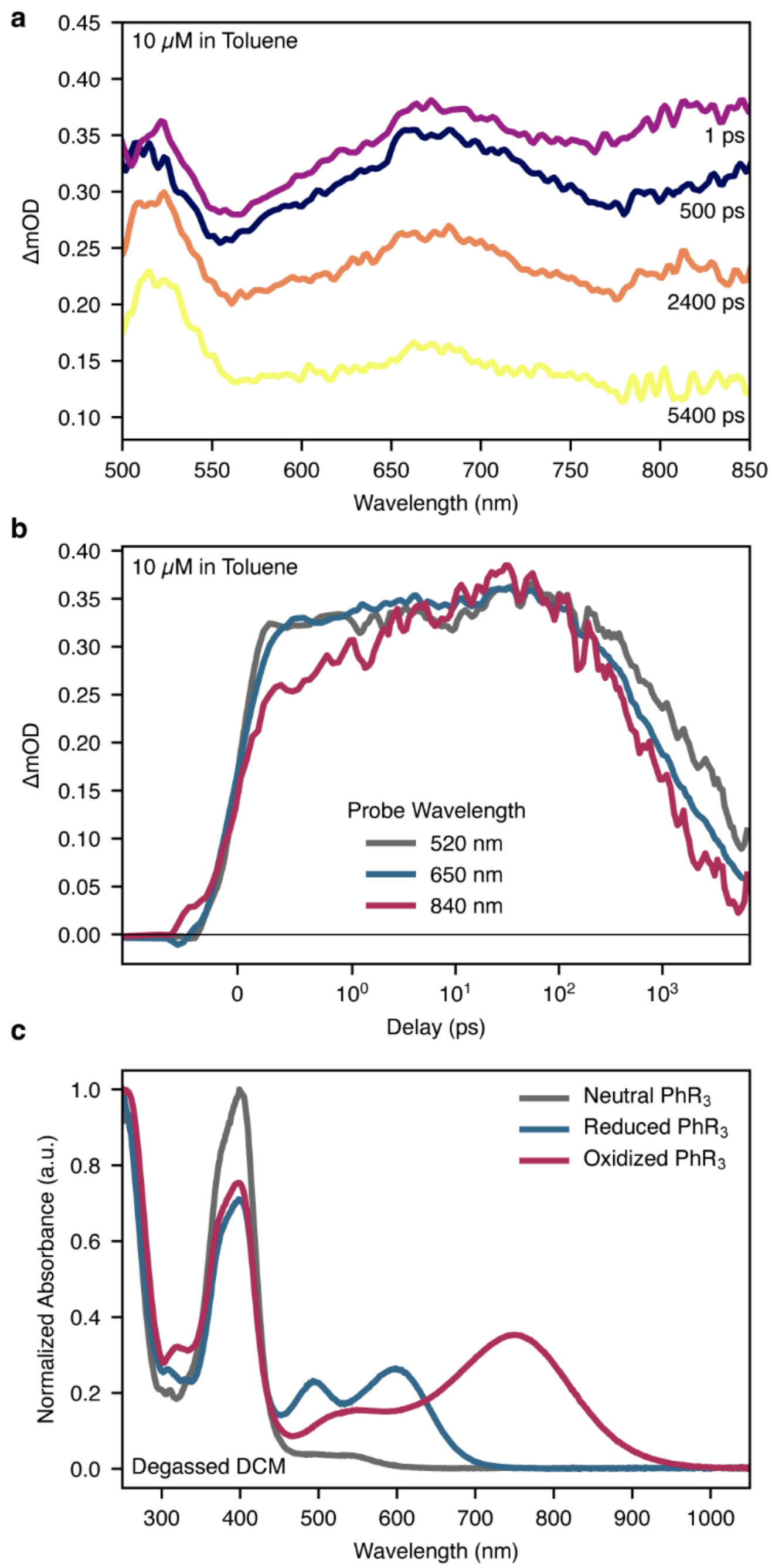

**Figure 2. Probing the electronic structures of the excited states.** Transient absorption on a 10 μM toluene solution of $PhR_3$ with **(a)** spectra at select times and **(b)** kinetic traces at select wavelengths. Sample excited with 380 nm light at ~50 μJ/pulse. **(c)** Spectroelectrochemistry of $PhR_3$ in degassed DCM, showing absorption spectra of neutral, reduced, and oxidized $PhR_3$. The concentration of the neutral triradical was adjusted to an optical density of 1.0.

Magnetic characterization of $PhR_3$ is presented in Fig. 3. SQUID magnetometry on a 5.9 mg bulk powder sample at 7T shows increasing magnetization with decreasing temperature attributed to a ferromagnetic ground state (Fig. 3a). Saturation of the magnetization around 3K is due to spin polarization into the $m_s$ = -3/2 sublevel of the ground state quartet by the 7T magnetic field following Boltzmann statistics. We fit the magnetization curve to a model composed of thermally-weighted populations of a Curie-like doublet and quartet to find a ferromagnetic ground state with a quartet-doublet energy gap of $\Delta E_{QD}$ = -11.5 ± 0.4K. We do not need to invoke intermolecular exchange to fit the magnetization data. Note that the SQUID measurement provides adequate precision to determine this small exchange energy, see Fig. S.II.4. As the ground state doublet and quartet states are both fourfold degenerate (as a consequence of $2^3 = 8$ states in the product basis), the low exchange energy results in an equal population of doublets and quartets at room temperature.

Continuous-wave electron paramagnetic resonance spectroscopy (cwEPR) on $PhR_3$ in a solid polystyrene matrix at X-band (~9.4 GHz) confirms the existence of both doublets and quartets at room temperature (Fig. 3b). Fitting the cwEPR spectrum using a model Hamiltonian composed of a spin triangle (see SI Section III) yields a pairwise dipolar coupling strength of $D$ = 26.4 MHz where the quartet $m_s$ = ±3/2 sublevels are split from the ±1/2 sublevels by an energy gap of 3D/2. Note that the pairwise coupling strength gives axial and rhombic dipolar couplings of 19.8 MHz and 0 MHz (SI Section III). Including g-anisotropy with $g_\perp$ = 2.0024 and $g_\parallel$ = 2.0032 accounts for the slight asymmetry in the spectrum. The pairwise dipolar coupling of 26.4 MHz leads to zero-field splitting between the $m_s$ = ±1/2 and ±3/2 sublevels of 39.6 MHz, which is in excellent agreement with the zero-field splitting of previously reported meta-phenyl linked TTM diradicals.[16,26] This agreement is expected due to the preservation of the alternant hydrocarbon structure when adding the additional radical, meaning the spin-spin coupling should not change significantly from the diradical case. We note that none of the magnetic characterization and its modeling discussed throughout shows any evidence of a triplet impurity which would arise from incomplete deprotonation.

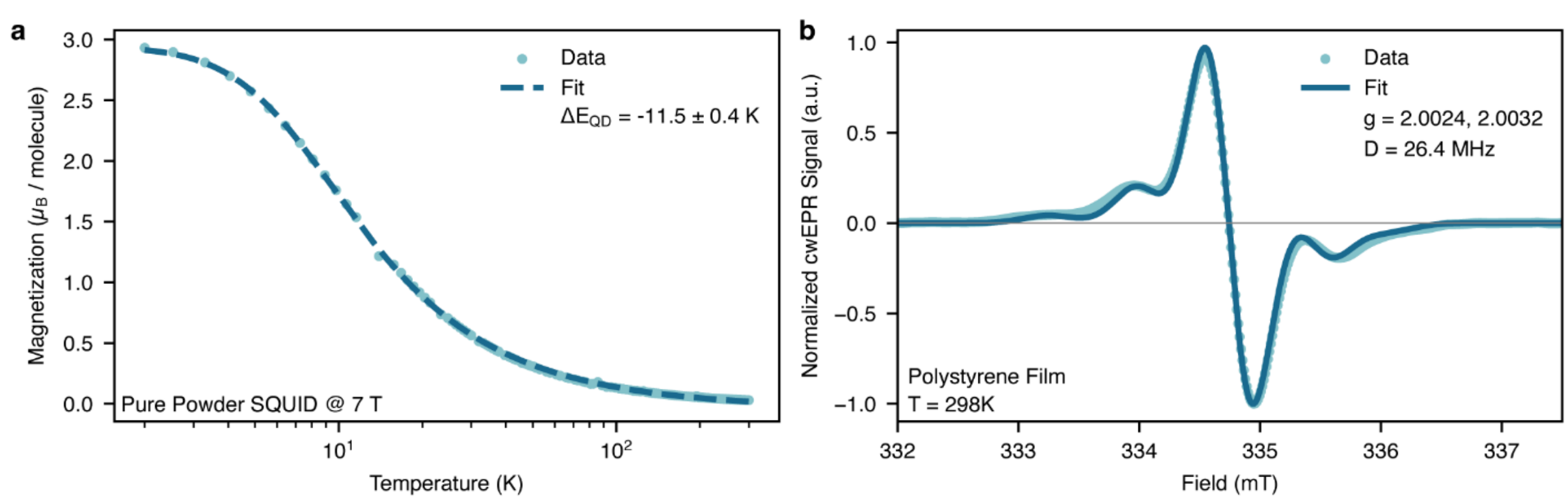

Figure 3. Ground-state magnetic characterization of the triradical. **(a)** Pure powder SQUID magnetometry on $PhR_3$. The magnetization is fitted to a model incorporating a Curie-like doublet and quartet which are separated by an exchange energy. **(b)** X-band EPR spectrum of $PhR_3$ in a polystyrene matrix. The data is fit to a spin Hamiltonian describing three electrons distributed on an equilateral triangle and the z-direction pointing along the principal axis.

**Optical spin polarization of the quartet ground state.** The prompt (t = 1 $\mu$s) transient EPR spectrum of $PhR_3$ in a solid polystyrene matrix at 100K following a 410 nm laser flash is shown in **Fig. 4a**. We observe a non-Boltzmann spin polarization that is long-lived with a lifetime of 37.4 ± 0.1 $\mu$s (**Fig. 4b**), with no temporal change in the lineshape (**Fig. S.II.5**). The ~$\mu$s lifetime for the spin-polarization compared to the ~ns excited state lifetime requires that the transient EPR signal arises from spin polarization in the ground state. Fitting the prompt spectrum in **Fig. 4a**,[44] we find overpopulation of the ±1/2 sublevels of the quartet based on the (*A*, *E*) pattern observed in the trEPR. The location of these transitions based on the fit is shown schematically in **Fig. 4c**. We note that the fit parameters for the trEPR experiment closely agree with the dark cwEPR spectrum in **Fig. 3b**, further indicating ground state spin polarization.

We propose that polarization arises due to spin-selective intersystem crossing (ISC) from the excited state quartet $Q_1$ to the doublet $D_2$, with sublevel-selective ISC from the $m_s = \pm 3/2$ quartet states. Considering the small exchange energy splitting between $Q_0$ and $D_0$ (measured from the SQUID magnetometry, **Fig. 3a**), we expect that the excited states $Q_1$ and $D_2$ are also separated by a small energy gap that is much less than thermal energy at room temperature. This small exchange enables fast ISC (on the nanosecond timescale, vide infra) from $Q_1$ to $D_2$, followed by ultrafast (picosecond timescale) relaxation to $D_1$ via internal conversion. The internal conversion process suppresses back transfer from $D_2$ to $Q_1$, enhancing spin polarization. [Note that $Q_1$ and $D_1$ are *not* separated by exchange as $D_1$ features a distinct zwitterionic electronic structure.]

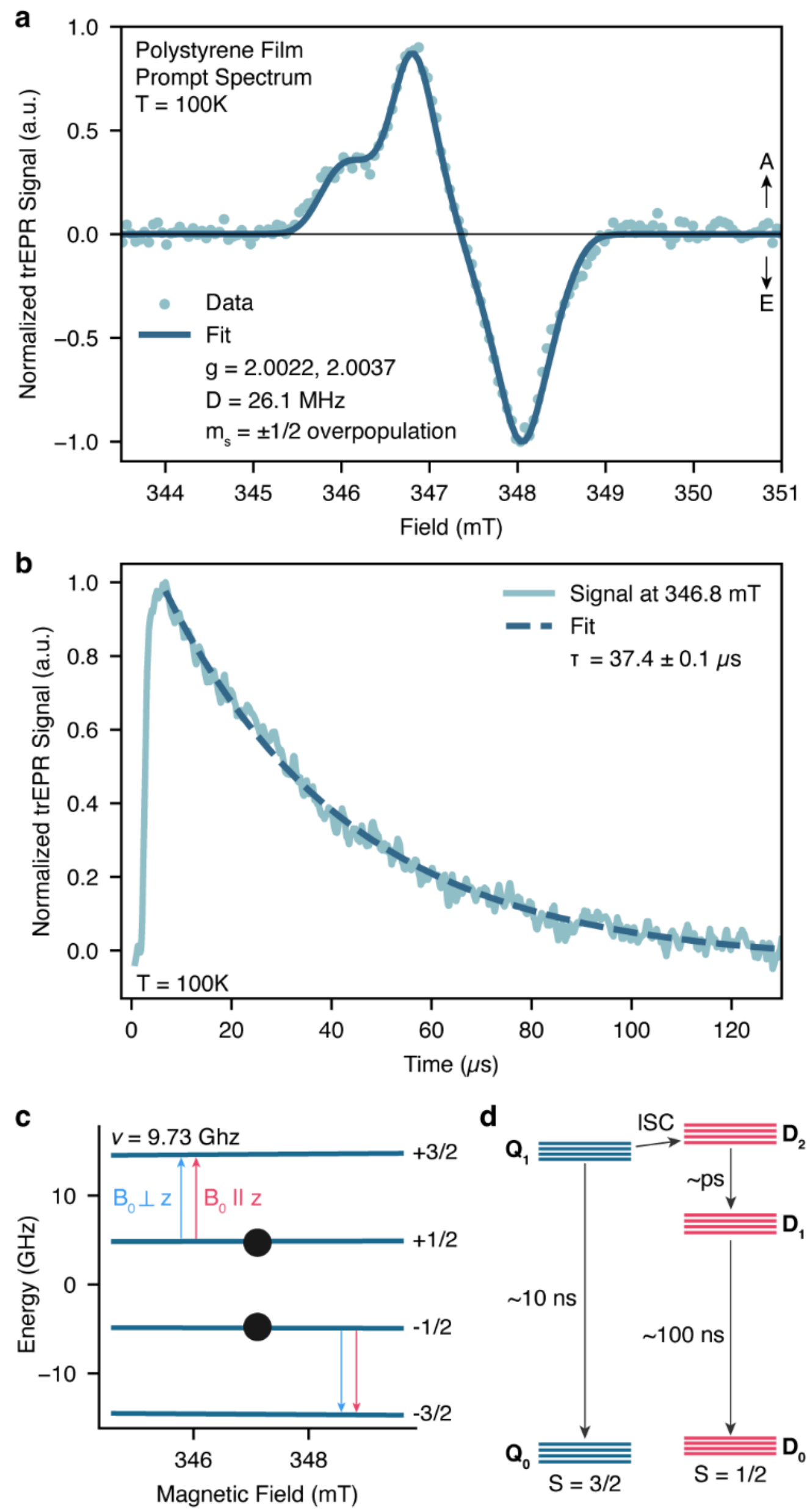

**Figure 4.** Optical ground-state spin polarization. **(a)** Prompt transient EPR spectrum at X-band on a $PhR_3$ in polystyrene film at 100K. The plotted spectrum is recorded 500 ns after photoexcitation by a 410 nm laser operating at 54 mW/cm$^2$ and 20 Hz. **(b)** Decay trace of the trEPR absorption signal at 346.8 mT in (a). The decay is fit to a monoexponential to yield a polarization lifetime of 37.4 ± 0.1 $\mu$s. **(c)** Zeeman-splitting diagram of the triradical using the fit parameters from (a). The trEPR transitions are labeled with an overall (*A*, *E*) pattern from spin-polarization into the $m_s$ = ±1/2 quartet sublevels. **(d)** Proposed mechanism for optical spin polarization. Upon photoexcitation, $Q_1$ undergoes fast ISC to $D_2$ which is selective for the ±3/2 sublevels. Ultrafast (~ps) internal conversion to the lower-lying $D_1$ state prevents back-transfer of spin population to the quartet manifold. The timescale for the PL is derived from TCSPC, while the internal conversion timescale is estimated from TA.

**Coherent spin control of the quartet ground state.** We demonstrate coherent control of the ground state quartet with pulse EPR at X-band (~9.7 GHz). We report pulse EPR on a 200 $\mu$M solution of $PhR_3$ in frozen toluene at 80K (**Fig. 5**); the same experiments performed in a polystyrene film are shown in **Fig. S.II.6**. As a control, we confirm that the echo-detected field sweep (EDFS) spectrum agrees with the cwEPR spectrum (**Fig. S.II.7**). We first examine the spin-lattice relaxation time ($T_1$) which sets the upper limit of the ground state spin polarization (**Fig. 5a**). Inversion recovery was performed at 347.5 mT with π and π/2 times of 32 and 16 ns and a recovery delay time of 300 ns. Fitting the echo intensity to a monoexponential, we find a $T_1$ of 157 ± 2 $\mu$s which is similar to previous reports on TTM diradicals.[15,16,26] Changing to a polystyrene matrix lengthens $T_1$ to 246 ± 10 $\mu$s at 80K, highlighting an opportunity to utilize $PhR_3$ for $T_1$ relaxometry based sensing.[4,25,45,46]

We now turn to the phase memory time ($T_m$), which is approximately equal to the coherence time $T_2$ (**Fig. 5b**). We use a two-pulse Hahn echo sequence with varying recovery times between pulses with the same π time as the inversion recovery experiment. Fitting the echo intensity to a stretched exponential model where intensity $\propto \exp[-(2\tau/T_m)^x]$,[22,47,48] we find a $T_m$ of 4.04 ± 0.03 $\mu$s. Both the spin-lattice and phase memory times are sufficiently long for quantum information applications, and we expect that further design of molecular structure and quantum control sequences such as dynamic decoupling to extend coherence times.

To operate the ground state quartet as a qudit, we need to separately address the outer quartet transitions, i.e. $m_s$ = +1/2 to +3/2 and -1/2 to -3/2. This capability is demonstrated in a transient nutation experiment using the phase-inverted echo-amplitude detected nutation (PEANUT) pulse sequence (**Fig. 5c**).[49] In PEANUT, we measure transient nutations at varying fields; applying a fast Fourier transform (FFT) reveals the nutation frequencies of the transitions at each field. The full transient nutation map shows peaks in the FFT centered at 18.2 and 24.9 MHz. We assign these peaks to the doublet and quartet transitions respectively as per their multiplicities. [Note that the frequencies of the doublet and quartet transitions do not follow the typical $\sqrt{S(S+1)-m_s(m_s+1)}$ scaling due to weak exchange coupling in the triradical as noted previously.][13,49–51] Taking slices of the map at specific frequencies shows the equivalent of the echo-detected field sweep (EDFS) spectrum of the ground state doublet and quartet (**Fig. 5d**). The multiplicity of the states is clearly observed, with a single peak for the doublet transition and three peaks for the quartet. The energetic separation of the quartet transitions may allow use of the $m_s$ = +1/2 to +3/2 and -1/2 to -3/2 transition as independent qubits, with the $m_s$ = -1/2 to +1/2 transition acting as a possible entangling operation.

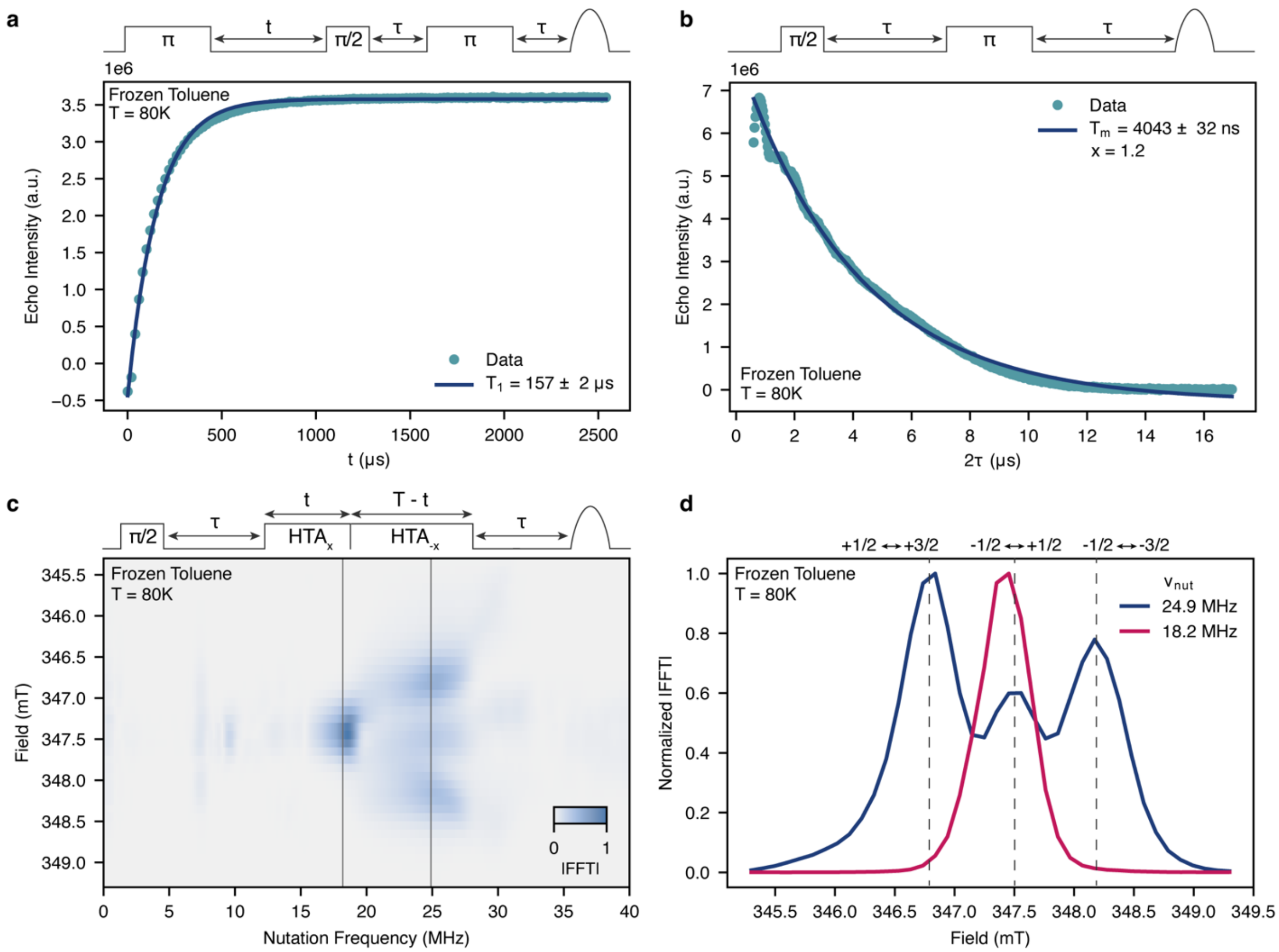

**Figure 5**. Coherent control of the quartet ground state. **(a)** Inversion recovery measurement of spin-lattice relaxation time $T_1$ at 347.5 mT. **(b)** Echo-detected measurement of the phase memory time $T_m$ at 347.5 mT. $T_m$ is fit with a stretched monoexponential model where x is the stretching coefficient. **(c)** Transient nutation map using the PEANUT sequence. In PEANUT, a two-part high turning angle (HTA) pulse is used in place of a standard nutation pulse used to coherently drive oscillations. The colormap shows the fast Fourier transform (FFT) of the nutations at each field. **(d)** Slices of the transient nutation map at 18.2 and 24.9 MHz. All measurements were performed at 80K in a glassy toluene matrix; see Supplementary Information for measurements in polystyrene film. The length of the π pulse is 32 ns.

**Room temperature optically detected magnetic resonance.** Having established optical initialization and ground state control of $PhR_3$, we now turn our attention to whether we can utilize the triradical for optically detected magnetic resonance (ODMR). The zero-field continuous wave ODMR spectrum of a $PhR_3$ film at room temperature is shown in Fig. 6a. We observe ΔPL/PL values up to $-1.7 \times 10^{-4}$ with a broad negative resonance at 106 MHz and a smaller, lower frequency feature around 35 MHz. Based on the ground-state *D*-value of 26.4 MHz measured from cwEPR (Fig. 3b), we consider the 106 MHz resonance to originate from manipulation of the excited state while the 35 MHz corresponds to a ground state transition. Fitting the spectrum with the same spin model used for the EPR yields ground- and excited-state dipolar couplings of *D* =

22 MHz and 73 MHz (**Fig. S.II.8**). Spectral filtering of the PL with a 650 nm short-pass filter enhances the negative ODMR contrast, while the use of an 800 nm long-pass filter results in a positive contrast. This sign inversion suggests that the resonant microwave field transfers population from the brighter excited state quartet (633 nm PL) to the dimmer doublet (690 nm PL) manifold. Based on the proposed spin polarization mechanism (**Fig. 4d**), the microwave field would transfer spin population from the $m_s = \pm 1/2$ to the $\pm 3/2$ sublevels of the $Q_1$ state, leading to more ISC to $D_2$ and subsequently enhanced $D_1$ luminescence. We note the sign inversion is more pronounced when filtering with a 700 nm long-pass filter at 77K (**Fig. S.II.9**) due to the suppression of the vibronic overtone of the $Q_1$ at lower temperature.

Zeeman splitting of the ODMR out to 20 mT and at 77K is shown in **Fig. 6b**, with room temperature data reported in the Supplementary Information (**Fig. S.II.10**). Above ~5 mT, there is one dominant feature that moves linearly with applied field, corresponding to Zeeman splitting of the excited state quartet. As with the frequency-resolved EDFS measurement (**Fig. 5d**), we can observe energetic separation of the outer quartet transitions ($m_s = +3/2$ to $+1/2$ and $-3/2$ to $-1/2$) set by the zero-field splitting, observed in **Fig. 6b** as two parallel tracks separated by around 100 MHz and seen more clearly in **Fig. S.II.11**, although powder averaging and broad linewidths prohibit cleaner separation of the transitions. At low fields (**Fig. 6b** inset) we observe a feature around 0 mT and 0 MHz, as $B_0 > 0$ lifts the degeneracy of the $m_s = \pm 1/2$ and $\pm 3/2$ pairs. Also visible is a faint line around 200 MHz and 2.5 mT corresponding to a $\Delta m_s = 2$ transition.

While the excited state appears to predominate these ODMR spectra, further developments in chemical design (e.g. crystalline samples providing narrower linewidths) and measurement conditions (e.g. increased optical excitation rate; time-resolved measurements) provide avenues for improved addressability and manipulation of the ground-state.

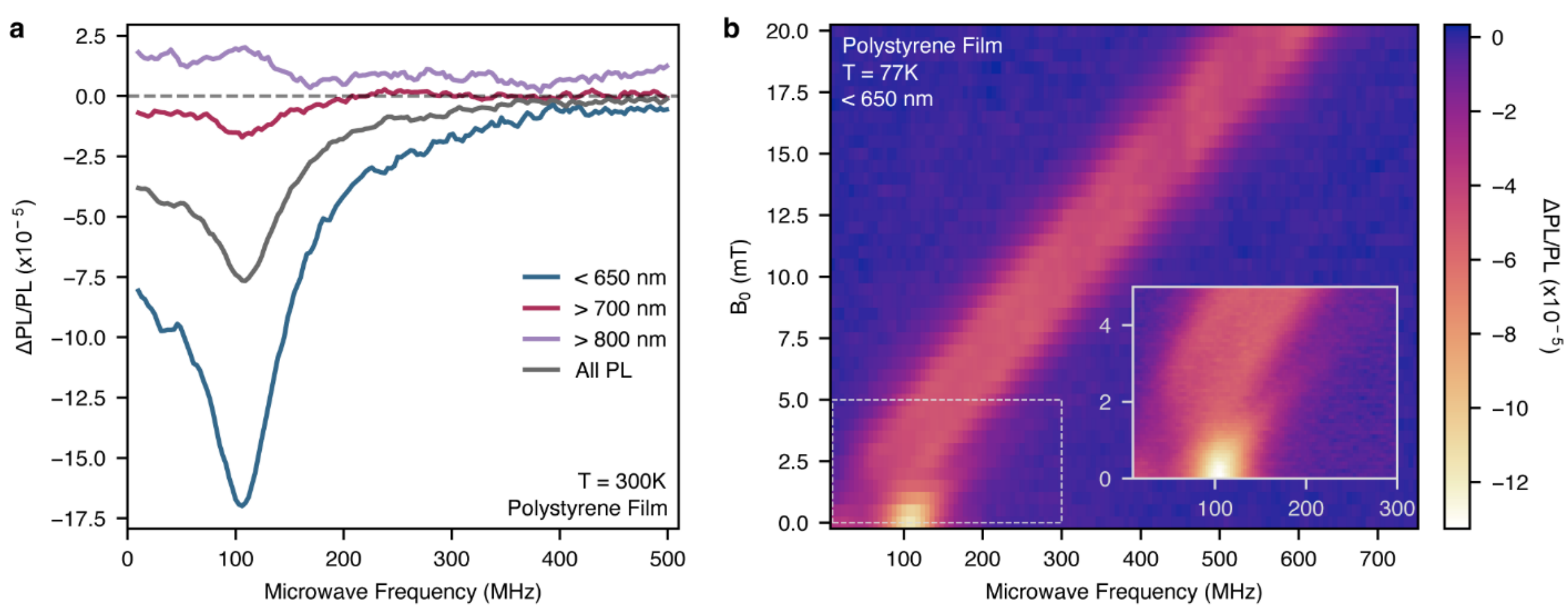

**Figure 6.** Optically detected magnetic resonance of the triradical. **(a)** Zero-field ODMR of a $PhR_3$ in polystyrene film under ambient conditions. The blue, red, and purple traces show spectrally filtered ODMR using <650 nm PL, >700 nm PL, and >800 nm PL. Measurements performed on a 1.25 wt% $PhR_3$ in polystyrene film under 405 nm excitation. **(b)** Zeeman splitting of the ODMR under an applied magnetic field at 77K. The <650 nm PL is collected to isolate the quartet signal. The inset shows a separate measurement around zero field with a finer field and frequency sweep.

**Magneto-photoluminescence.** The low dipolar coupling stemming from spatial separation of meta-phenyl linked diradicals has previously allowed sensing of magnetic fields on the order of a few mT.[15,16,26] For $PhR_3$, we find a magneto-photoluminescence (MPL) response of 1.5% within 5 mT when integrating over all PL (**Fig. 7a**). Spectrally filtering the PL using 650 nm short pass and long pass filters, we find that the majority of the MPL signal arises from the 633 nm band. Spectrally resolving the PL contrast indeed confirms that luminescence is suppressed in the 633 nm band upon applying a magnetic field (**Fig. 7b**), while we observe enhancement at higher wavelengths. This sign inversion with respect to wavelength matches the zero-field ODMR results (**Fig. 6a**) and indicates transfer of spin population from the excited quartet to the doublet under applied magnetic fields. Based on the linewidth of the MPL response, we ascribe this effect to mixing of the zero-field $Q_1$ sublevels by the magnetic field which impacts ISC rates (see "MPL modelling" in the **Supplementary Information**). Time-resolving the MPL at 10 mT shows a transient decrease in PL within 50 ns following photoexcitation (**Fig. S.II.12**). This response is faster than the 91.5 ns PL lifetime of the $D_1$ state, highlighting that ISC between the quartet and doublet manifolds is competitive with the PL from both manifolds.

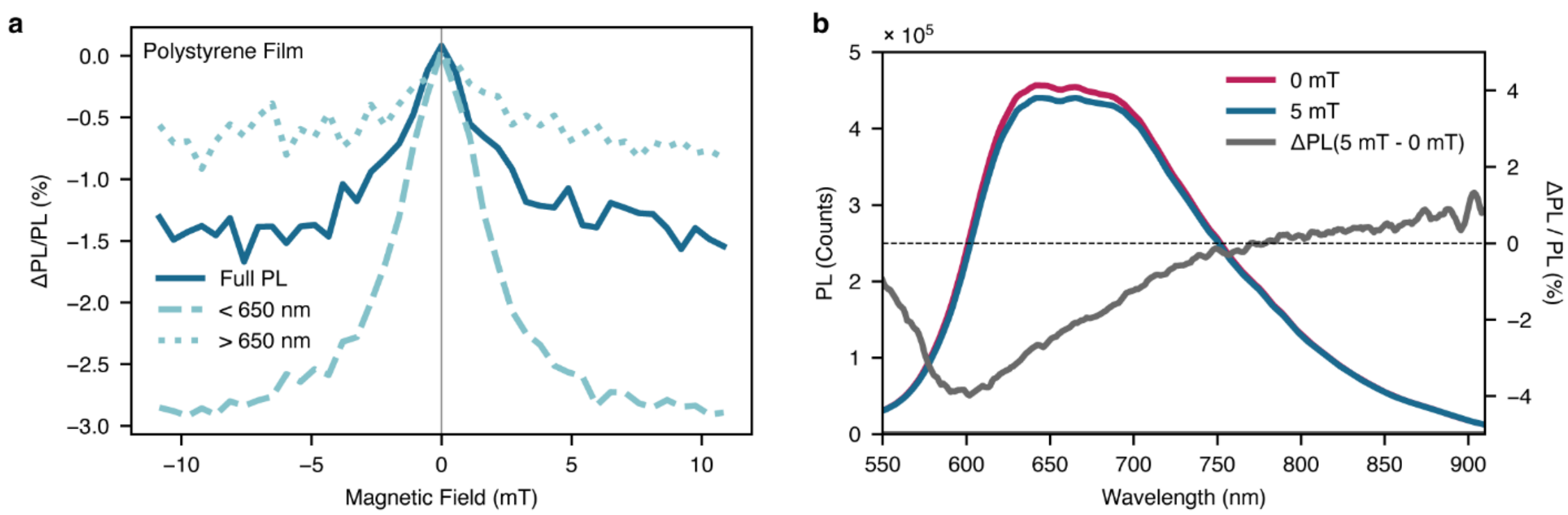

**Figure 7.** Magneto-photoluminescence of the triradical. **(a)** Magneto-photoluminescence of a 0.05 wt% $PhR_3$ in polystyrene film of the spectrally-integrated PL, <650 nm PL, and >650 nm PL. The film was excited with a 405 nm laser at 100 $\mu$W/cm$^2$. **(b)** Spectrally-resolved magneto-photoluminescence of the $PhR_3$ film at 5 mT. The film was excited with a continuous wave 405 nm laser (power density 100 $\mu$W/cm$^2$) and spectra were measured using an intensified camera with the magnet periodically switched on and off for 1s cycles to measure the change in PL.

## Discussion

We have established a luminescent triradical as an optically-addressable qudit with potential for optically detected quantum sensing. Compared to luminescent diradicals,[15,16,26] the additional radical center introduces a richer manifold of spin physics with a doublet and quartet manifold. The quartet introduces an additional spin state ($m_s$ = -3/2, -1/2, +1/2, +3/2) compared to triplets ($m_s$ = 0, ±1) that can be separately addressed. We anticipate the presence of a third spin to open up possibilities for quantum control. For example, this additional spin state could be used for on-qudit entanglement to improve sensitivity or in multi-quantum sensing protocols.[8,10,50]

We highlight the potential advantage of using ground states as a qudit rather than excited states to achieve longer state lifetimes without sacrificing optical readout efficacy. Excited states should have long lifetimes for quantum control, i.e. radiative rates $k_r$ should be small. This presents a trade-off for optical readout where we desire large $k_r$ for high PLQEs. The ground state qudit decouples the electronic states used in quantum control (ground state) and optical read out (excited state), removing this constraint entirely.

Finally, we note that the symmetry of the triradical here poses an interesting design opportunity for future generations of luminescent organic qudits. The pseudo-$C_3$ symmetry of the molecule forces exchange and dipolar interactions between radical centers to be equivalent. However, introducing controlled asymmetry may allow greater energetic separation of spin sublevels and improve our ability to selectively address desired states, currently being explored further.

## Methods

**Materials.** 2-(4-(Bis(3,5-dichloro-2',4',6'-trimethyl-[1,1'-biphenyl]-4-yl)methyl)-3,5-dichlorophenyl)-4,4,5,5-tetramethyl-1,3,2-dioxaborolane ($M_2$TTM-Bpin) was synthesized from tris(2,4,6-trichlorophenyl)methane (αHTTM) based on the previously published synthetic protocol.[37] All other reagents, catalysts and (anhydrous) solvents were purchased from Merck, BLDpharm, Fisher Scientific and TCI and used as received. Deuterated solvents were obtained from Eurisotop.

**Chemical Characterization.** NMR spectra were recorded on a 400 MHz Avance NEO 400 spectrometer ($^1$H, 400 MHz; $^{13}$C, 101 MHz). Chemical shifts are reported in δ (ppm) relative to the solvent peak: chloroform-*d* ($CDCl_3$: $^1$H, 7.26 ppm; $^{13}$C, 77.16 ppm) and dichloromethane-$d_2$ ($CD_2Cl_2$: $^1$H, 5.32 ppm; $^{13}$C, 53.84 ppm). Mass spectra were recorded on a Bruker UltrafleXtreme MALDI TOF/TOF mass spectrometer. Flash chromatography was carried out using CombiFlash NextGen 300+ system and RediSep silica gel columns.

**Electrochemistry.** Cyclic voltammograms (CV) were obtained at room temperature using an PalmSens EmStat4X potentiostat. The experiment was performed in a cylindrical 2.5 cm x 5 cm glass cell. A 3 mm diameter glassy carbon disc electrode was used as a working electrode which was polished carefully with 0.05-micron Al particle paste on a polishing pad, a Pt wire was used as a counter electrode, and as a reference electrode a pseudo reference $Ag/Ag^+$ silver wire was used. The pseudo reference electrode was calibrated against ferrocene/ferrocenium ($Fc/Fc^+$) redox couple (the $Fc/Fc^+$ half-wave potential was determined at 0.31 V vs. $Ag/Ag^+$). As a supporting electrolyte 0.1 M tetrabutylammonium hexafluorophosphate ($nBu_4PF_6$) was used in anhydrous DCM, and the scan rate for CV was 0.1 V $s^{-1}$ while for measuring differential pulse voltammetry (DPV) the scan rate was 0.05 V $s^{-1}$. The electrolyte was bubbled with Argon before each measurement to remove any dissolved oxygen. Sample concentration was in the order of $10^{-5}$ M.

**Photoluminescence Quantum Efficiency.** PLQE measurements were acquired using a custom-built PLQE set-up. Samples were placed inside an integrating sphere and photoexcited from the substrate side using a 405-nm continuous-wave laser excitation source (Thorlabs) (laser power = 2 mW) with a spot size of 0.15 $cm^2$. The PL was collected using a fiber bundle (Ocean Optics QR600 7 SR125BX) coupled with a spectrometer (QE Pro, Ocean Optics). The PLQE was extracted from the photon energy (hf) and photon numbers of the excitation and emission obtained from numerical integration using Python. Three different spots were measured on the substrate.

**Spectroelectrochemistry**. A commercially available PalmSens EmStat4S potentiostat was connected to an Agilent Cary 60 UV-Vis spectrophotometer. The measurements were carried out in a custom-made three-electrode setup using a quartz cuvette (0.1 mm path length) as the spectroelectrochemical (SEC) cell, a platinum mesh as the working electrode (in the light path), a platinum wire as the counter electrode and a freshly activated silver wire as the Ag/Ag+ reference electrode. The silver wire was activated by immersing in a concentrated HCl solution to remove any silver oxides or other impurities, then rinsed with water and acetone and dried before measurements. The reference electrode was calibrated against Fc/Fc+ redox couple. The supporting electrolyte was 0.1 M solution of $nBu_4PF_6$ in anhydrous dichloromethane. The electrolyte was bubbled with Ar gas before measurements to remove any dissolved oxygen. The absorption spectra of the oxidized (cationic) and reduced (anionic) species were measured by increasing the applied potential until +1.5 V and –2.0 V versus Fc and Fc+, respectively. These are roughly 0.8 V higher than the corresponding half-wave potentials determined for the parent triradical to confirm the absence of additional redox events.

**Steady-state optical absorption and PL spectroscopy.** Steady-state optical absorption was measured a commercially available Shimadzu UV-2550 spectrophotometer. Steady-state PL spectra were measured using the home-built transient PL setup described below with the electrically gated intensified CCD (iCCD) camera operating in cw mode. Films were excited using laser pump pulses generated from an Orpheus-Lyra (Light Conversion). All films were prepared to be 0.05 wt% in polystyrene.

**Time-resolved PL spectroscopy.** Time-resolved PL spectra were measured using an electronically gated intensified charge-coupled device (iCCD) camera (Andor iStar DH740 CCI-010) connected to a calibrated grating spectrometer (Andor SR-303i). Pump laser pulses were generated from an Orpheus-Lyra (Light Conversion). Temporal evolution of the emission was obtained by stepping the intensified charge-coupled device delay with respect to the excitation pulse, with a minimum gate width of 5 ns.

**Transient absorption spectroscopy.** Picosecond transient absorption spectroscopy was measured using a commercial HARPIA (Light Conversion) system. The fundamental laser beam of the HARPIA at 1,030 nm was provided by PHAROS (Light Conversion). The ORPHEUS (Light Conversion) system, which was pumped by an ytterbium-doped solid-state chirped pulse amplifier, generated pump lights using 90% of the fundamental laser. The pump wavelength can be varied from 350 nm to 2,000 nm with a 10-kHz repetition rate and an ~100-fs pulse duration. The rest of the 10% fundamental beam was injected into HARPIA-TA to generate the probe light. The pump–probe time delay was provided by a mechanical delay stage. The probe pulses were detected by a spectrograph with dual outputs to cover the UV–near-infrared regions (Kymera-193i-B2).

**SQUID magnetometry.** Powder samples were encapsulated and mounted in a diamagnetic gelatin capsule and inserted into a straw sample holder. The ends were held in place using non-magnetic tape, with the sample position being ~66 mm in the sample area within the magnetometer. The DC-magnetization was measured by warming on a Quantum Design MPMS3 in the temperature range 2.0 – 300 K under an applied magnetic field of 70000 Oe (7T), after cooling the sample from 300 K to the base-temperature (2-4 K) in zero field (ZFC).

**X-band electron paramagnetic resonance spectroscopy.** EPR samples were prepared to an approximate concentration of 200 $\mu$mol/dm$^3$ in a solid polystyrene matrix. Polystyrene films were deposited into quartz EPR tubes from a toluene/polystyrene precursor that was dried under vacuum and subsequently flame sealed. X-band EPR spectroscopy was measured in the Centre for Advanced ESR (CAESR) in the Department of Chemistry of the University of Oxford.

For continuous wave measurements, a Bruker EMXmicro X-band spectrometer with a PremiumX bridge was used with a Bruker ER4122SHQE-W1 resonator. Non-saturating conditions were found with 9.5 $\mu$W microwave power and a 100 kHz field modulation of 4 G. Field calibration from the Hall sensor was performed with N@C60 using a g-value of 2.00204(4).[51]

For laser-induced transient and pulsed EPR measurements, a Bruker EleXSys I-DU E680 spectrometer with an E580 X-band bridge as used with a Bruker ER4112X-MD5W1 resonator. Non-saturating conditions were found with 5 $\mu$W microwave power. Temperature control was provided by an Oxford Instruments CF935O cryostat using an ITC-503S controller. Excitation was provided at 410 nm by an Opotek Opolette HE355 OPO laser set at 20 Hz and 54 mW / cm$^2$ optical power density, with time synchronization provided by a Stanford Research Systems DG645 delay generator.

**Continuous-Wave Optically Detected Magnetic Resonance (cw-ODMR).** For field dependent cw-ODMR, samples were excited using a 405 nm laser diode (ThorLabs LP405-SF10), which was filtered using a 450 nm short-pass filter (ThorLabs FESH0450). A dichroic mirror (ThorLabs DMLP425R) was used to direct the excitation laser to a 1” achromatic doublet lens (ThorLabs AC254-030-AB) which was used to focus the light into a 400-micron diameter core optical fiber (FT400EMT) that was placed in contact with the sample. The photoluminescence was collected by the same fiber, then collimated via the lens, separated from the laser excitation via the dichroic, and passed through a 550 nm long-pass filter (ThorLabs FELH0550) before being focused onto a free-space detector (FEMTO, OE-200-SI) using a second achromatic doublet (ThorLabs AC254-030-AB). For spectrally selective measurements, different filters were added to the collection path. For 77K measurements, the sample was submerged in liquid nitrogen in a dewar.

Zero-field measurements, room temperature measurements were performed as above, but without the optical fiber in contact with the sample. In this case, the excitation laser was focused onto, and PL collected from the sample.

The microwave fields were applied using a co-planar waveguide connected to a signal generator (Rohde & Schwarz SMA100B) with a modulation rate of 117 Hz, and amplified using a high-power microwave amplifier (Mini-circuits LZY-22+ or ZHL-5W-202-S+). The PL signal from the free-space detector was passed into a lock-in amplifier (Stanford Research Systems, SR830) which was referenced to the microwave modulation frequency. The lock-in signal yielded the ODMR signal $\Delta$PL= PL(Microwaves on) - PL(Microwaves off) which we normalize to the DC PL which we measure using an auxiliary input to the lock-in amplifier.

To apply a static magnetic field $B_0$, we used a home-built solenoid mounted to the probe. The magnetic field is oriented parallel to the optical axis, normal to the plane of the film. The magnetic field strength was calibrated by performing ODMR on an NV center diamond spin ensemble.

**Magneto-photoluminescence.** Magneto-photoluminescence spectra were collected using a home-built setup with a custom magnetic coil. The magnetic field was delivered using a copper wire wound ~1500 times around a hollow cylinder with a 2 mm diameter. Precise currents were supplied to the coil using a Keithley 2400 Source Measure Unit and the magnetic field generated per amp of supplied current was calibrated using a Hall probe. Samples were excited with a continuous wave 405 nm laser (ThorLabs LP405C1) that was chopped using an acousto-optic modulator (Isomet M1250-T250L-0.45). All spectra were measured with a 405 nm pump at 100 $\mu$W/cm$^2$.

For field-swept spectra, PL at each magnetic field was detected with a silicon avalanche detector (ThorLabs APD410A/M) and an SR830 lock-in amplifier (Stanford Instruments) monitoring at the laser modulation frequency. Zero-field reference measurements were collected before and after each applied field, with the reference PL value being calculated as the average of the two values to account for possible photobleaching.

For spectrally resolved magneto-photoluminescence, PL spectra were measured at 0 and 5 mT using an intensified CCD camera (Teledyne PI-MAX 4 2048f) operating in cw mode and an imaging spectrograph (Teledyne SpectraPro 2150i). PL from the sample was focused into the spectrograph using an achromatic doublet (ThorLabs AC127-050-AB) to correct for chromatic aberration. PL spectra were collected sequentially in 1 second acquisitions with the magnet switched on and off. To account for photobleaching, the difference spectrum was calculated using each 5 mT spectrum compared to the zero-field PL spectrum that preceded and followed it.

**Time-correlated single-photon counting and transient magneto-photoluminescence.** Time-correlated single photon counting was measured on a home-built setup using a 405 nm pulse laser (ThorLabs NPL41B). The sample was held in the magnet coil described above. PL was detected using a single photon avalanche diode (Micro Photon Devices PDM series) connected to a PicoHarp 300 for photon counting. A laser repetition rate of 5 MHz (5 $\mu$s period) was used for all measurements to avoid wrap-around effects. Transient MPL was measured by periodically switching the magnet on and off, collecting ~60s of PL with the field on and off. The time-resolved MPL was then calculated by comparing the TCSPC traces with the magnetic field on and off.

**Acknowledgements**

We thank Rituparno Chowdhury for his involvement at an early stage in the project. We thank Lujo Matasovic and Sebastian Gorgon for support and helpful discussions. We thank Shaoni Kar for assisting with the PLQE measurement. **K.K** acknowledges support from the Winston Churchill Foundation of the USA. **A.B.** and **P.M.** acknowledge funding from the Research Council of Finland (grant no. 363345). **S.B** and **A.I** acknowledge support by UK Research and Innovation (grant number MR/W006928/1). **W.K.M.** acknowledges funding from the John Fell Fund, grants nos. 0007019 and 0010710. **R.H.F** and **K.K** received funding from the European Research Council under the European Union's Horizon 2020 research and innovation programme (Grant Agreement No. BRIGHTS – 101200453).

**Data availability**

The data underlying this work are available at [URL to be added in proof].

## For Table of Contents Only

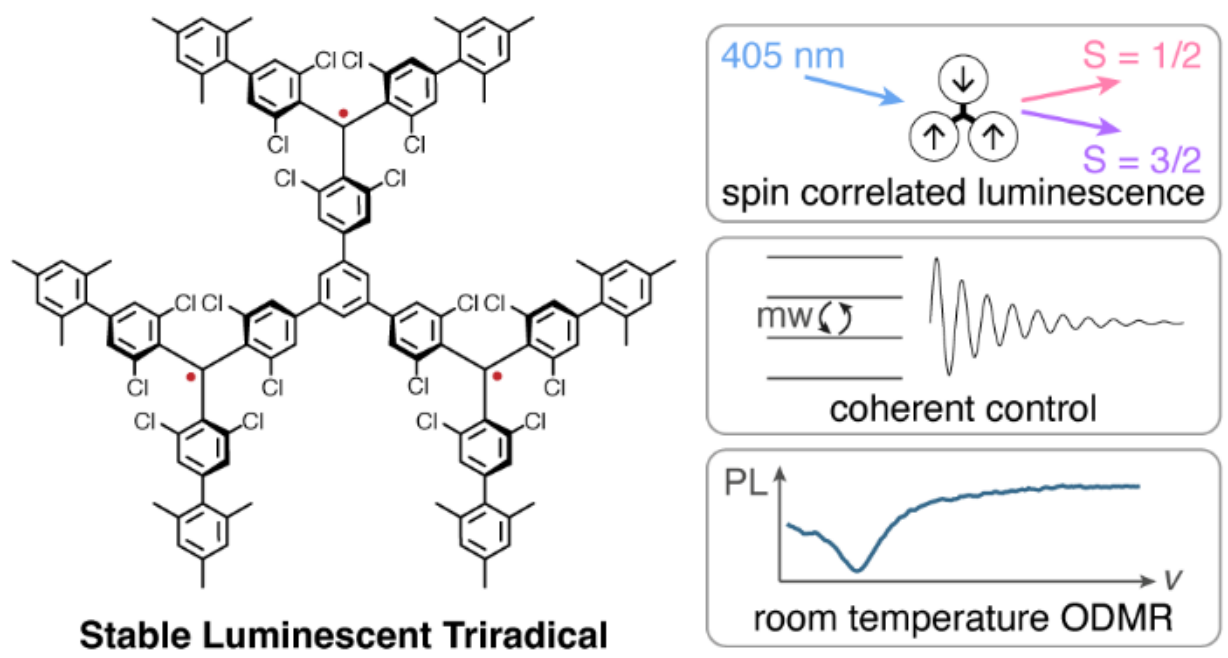